\documentclass[letterpaper,journal]{IEEEtran}

\usepackage{amsmath,amsfonts}

\usepackage{algorithm}
\usepackage{algpseudocode}

\usepackage{array}
\usepackage{booktabs}
\usepackage{multirow}
\usepackage{tabularx}
\usepackage{threeparttable}
\usepackage{siunitx}
\usepackage{graphicx}
\graphicspath{{./figs/}}
\usepackage[caption=false,font=normalsize,labelfont=sf,textfont=sf]{subfig}
\usepackage{stfloats}

\usepackage{tikz}
\usepackage[most]{tcolorbox}
\usepackage{xcolor}

\usepackage{textcomp}
\usepackage{paralist}
\usepackage{xspace}
\usepackage{cite}
\usepackage{url}
\usepackage{xurl}
\usepackage{verbatim}

\usepackage[hidelinks]{hyperref}

\ExplSyntaxOn

\NewDocumentCommand{\modelname}{m}
{
    \group_begin:
    \ttfamily
    \tl_set:Nn \l_tmpa_tl {#1}
    \tl_replace_all:Nnn \l_tmpa_tl {-} {-\allowbreak}
    \tl_use:N \l_tmpa_tl
    \group_end:
}

\ExplSyntaxOff

\definecolor{rqboxbg}{RGB}{246,247,249}
\definecolor{rqboxframe}{RGB}{120,126,135}
\definecolor{rqboxtitlebg}{RGB}{232,235,239}

\newtcolorbox{rqanswer}[1]{
    enhanced,
    breakable,
    colback=rqboxbg,
    colframe=rqboxframe,
    colbacktitle=rqboxtitlebg,
    coltitle=black,
    boxrule=0.6pt,
    arc=2mm,
    left=1.2mm,
    right=1.2mm,
    top=1.0mm,
    bottom=1.0mm,
    before skip=6pt,
    after skip=6pt,
    fonttitle=\bfseries,
    title={#1}
}

\newcommand{\approach}{\texttt{CoSTAR}\xspace}

\usepackage{enumitem}

\newcommand{\circlenum}[1]{\textcircled{\raisebox{-0.3pt}{\scriptsize #1}}}

\begin{document}

\title{CoSTAR: Data Synthesis-Driven Constraint-Aware COBOL Section Summarization for Legacy System Modernization}

\author{
    Hao~Lin,
    He~Jiang,
    Xiaochen~Li,
    Weihong~Sun,
    Yufu~Wang,
    Zhilei~Ren,
    and~Ang~Jia%
    \thanks{H. Lin, H. Jiang, X. Li, Y. Wang, Z. Ren, and A. Jia are with the School of Software, Dalian University of Technology, Dalian, China. H. Jiang is also with the DUT Artificial Intelligence Institute, Dalian, China. W. Sun is with Hi-Think Technology, Corp., Dalian, China.}%
    \thanks{Corresponding author: He Jiang. E-mail: jianghe@dlut.edu.cn.}%
}

\maketitle

\begin{abstract}
COBOL remains critical to governments, financial institutions, and large enterprises; yet, aging technologies, shrinking expertise, and missing documentation make modernization of COBOL-based legacy systems increasingly urgent. Before migration, code summarization is a common practice to support legacy system understanding. However, COBOL code summarization, especially on section-level, faces two key challenges: data scarcity and migration constraint preservation. To address these challenges, we propose \approach, an integrated framework that combines execution-validated data synthesis with constraint-aware model training. \approach repurposes general-purpose programming tasks to synthesize execution-validated COBOL code-summary data through LLM-based generation to overcome data scarcity. Based on the synthesized data, \approach augments target sections with relevant data declarations and natural-language explanations, and uses constraint-guided structured rationales to train smaller base LLMs. The trained LLMs preserve the migration constraints for COBOL section summarization. We evaluate \approach on both public and confidential enterprise COBOL systems. \approach effectively synthesizes 3,764 execution-validated training instances. Based on these instances, \approach built on 7B/8B base LLMs can improve these LLMs with average relative gains of 25.38\% on ROUGE-L, 53.84\% on METEOR, and 37.22\% on chrF. In real-world enterprise evaluation, \approach built on only \modelname{Qwen3-8B}, outperforms the enterprise-deployed \modelname{Qwen3-235B} in accuracy, completeness, and conciseness. These results show that \approach enables small, locally deployable LLMs to achieve performance competitive with substantially larger LLMs for privacy-sensitive COBOL legacy systems.
\end{abstract}


\section{Introduction}
\label{sec:introduction}

For over six decades, COBOL has supported critical services across governments, financial
institutions, and more than 40,000 enterprises~\cite{upadhaya2023understanding,agarwal2024tutorial}. Over 800 billion lines of COBOL code remain in active use; these systems process 80\% of financial transactions, 95\% of ATM transactions, and approximately USD~3 trillion in daily commercial transactions~\cite{opentext2023microfocus,dau2024xmainframe,taulli2020cobol}. However, the foundations sustaining these systems are becoming increasingly fragile. Decades of evolution have created intricate internal dependencies and incomplete or even misleading documentation~\cite{khadka2014professionals,assunccao2025contemporary}. Even worse, the pool of experienced COBOL developers is shrinking dramatically, while technical support is also receding~\cite{gangula2025comparative,strobl2020towards}. As reported, nearly 40\% of the most critical U.S. federal legacy systems relied on unsupported hardware or software, while 64\% operated with known cybersecurity vulnerabilities~\cite{gao2025critical_legacy}.
For organizations that still depend on these systems, modernization is becoming increasingly urgent. 

Legacy system modernization aims to improve the maintainability and evolvability of aging systems, typically requiring their business knowledge to be recovered and documented before migration~\cite{assunccao2025contemporary}. A seemingly attractive shortcut is direct migration, which bypasses this understanding process by translating legacy code directly into a modern language. However, this direction is impractically. As reported, DOGE planned to use AI to migrate over 60 million lines of COBOL at the U.S. Social Security Administration (SSA) within months; but after nearly a year of effort, the initiative ultimately failed~\cite{wired2025doge_ssa,connolly2025ssa_doge,ssa2027_congressional_justification}. This is unsurprising. On the one hand, traditional rule-based translation often carries accumulated technical debt into the target language, producing verbose and difficult-to-maintain ``JOBOL'' code~\cite{bloomberg2022jobol,aws2025bluage}. On the other hand, LLM-based translation is likewise unreliable. Even on real-world projects in data-rich languages such as Java and Python, the best evaluated LLM succeeded on only 8.1\% of translations. Worse still, human reviewers may overlook a substantial fraction of bugs in LLM-generated code~\cite{kabir2024stack}. Therefore, direct LLM translation cannot be trusted for mission-critical legacy systems, where behavioral deviations are unacceptable~\cite{resqsoft_translation_not_modernization,ibrahimzada2025alphatrans}.

Therefore, prior studies suggest that a more practical route is to understand legacy systems before migration~\cite{khadka2014professionals,swimm_accelerating_cobol_modernization}. Code summarization supports this process by distilling program behavior, data transformations, and business rules into concise natural-language descriptions for subsequent code migration and data transformation~\cite{wolfart2021modernizing, diggs2024leveraging}. Despite substantial progress, code summarization studies remain concentrated on mainstream languages such as Java and Python, leaving legacy languages (e.g., COBOL) comparatively underexplored~\cite{ahmed2022multilingual,zhang2024review}. However, applying existing techniques on COBOL code summarization pose two key challenges.

\begin{figure}[t]
    \centering
    \includegraphics[width=0.68\columnwidth]{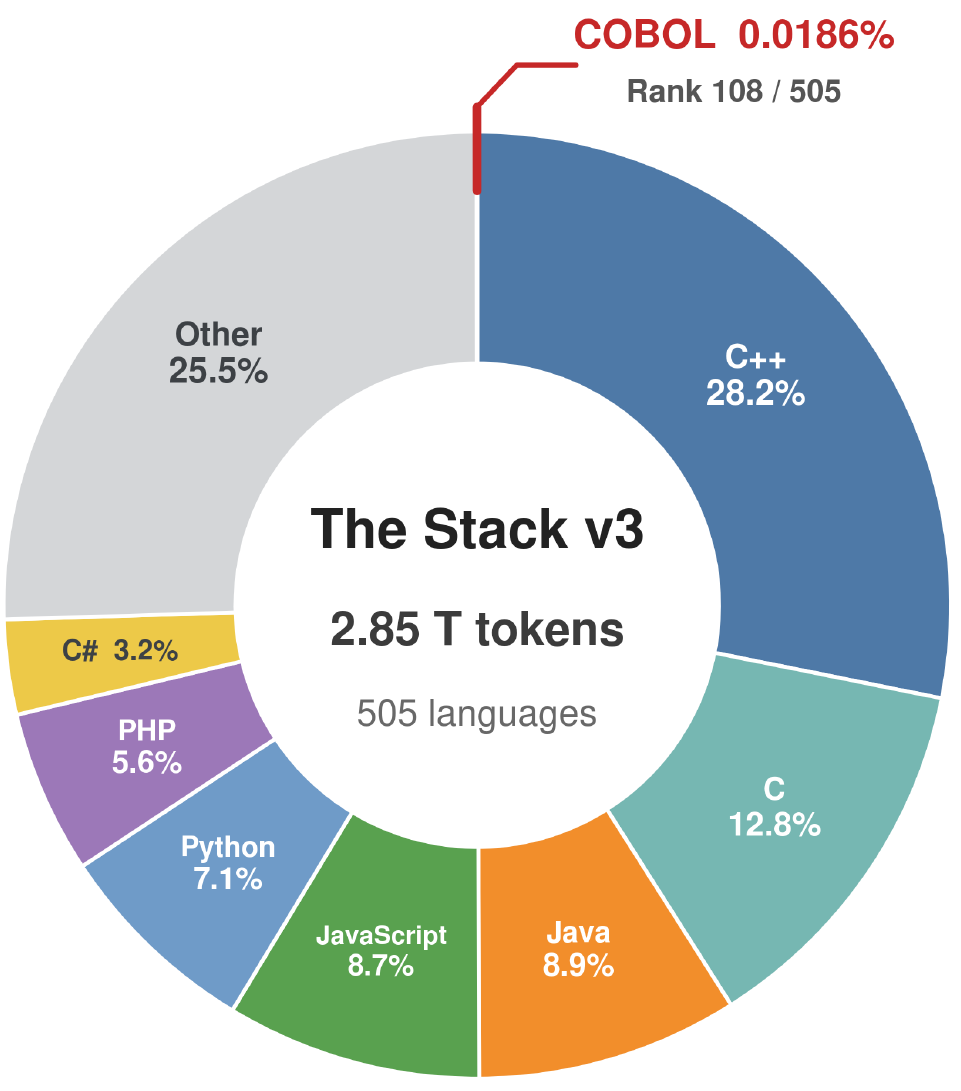}
    \caption{Programming-language token distribution in The Stack v3.}
    \label{fig:data_scarcity}
\end{figure}

\noindent\textbf{Challenge 1: Data scarcity.}
Existing LLM-based code summarization methods rely heavily on large-scale aligned code-summary training data~\cite{ahmed2022multilingual,zhang2024review}. Despite its extensive use in mission-critical sectors (e.g., government and finance), as statistics in \figurename~\ref{fig:data_scarcity}, COBOL accounts for only 0.0186\% of programming-language tokens~\cite{lozhkov2026stack-v3},
since much production COBOL code remains confidential. 
Moreover, the limited public-available COBOL code lack human-written summary. Most projects contain only file level description. In COBOL, \textit{section} is the natural functional unit for understanding business logic~\cite{ibm_cobol_program_format}, yet section-level summaries are practically absent from public data. This scarcity not only hinders developers from understanding section-level code, but also makes training summarization models at this granularity difficult.

\begin{figure}[t]
    \centering
    \includegraphics[width=0.92\columnwidth]{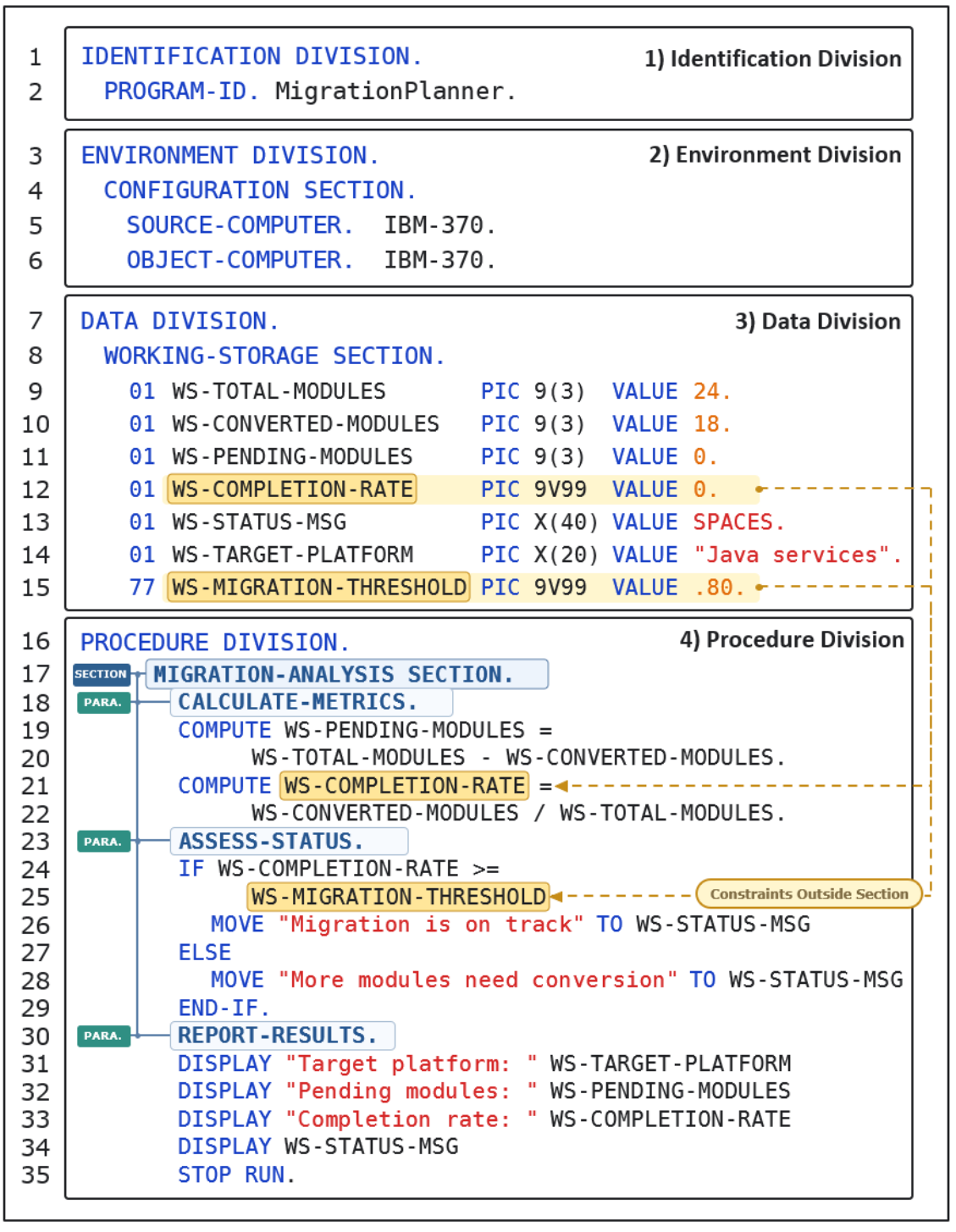}
    \caption{An example of COBOL code. COBOL separates data definitions from executable logic. The target section comprises multiple paragraphs but depends on data constraints defined separately in the data division.}
    \label{fig:data_logic_separation}
\end{figure}

\noindent\textbf{Challenge 2:  Migration constraint preservation.}
Modernization-oriented summaries should not only describe section behavior but also preserve the constraints needed to reproduce that behavior after migration. This is difficult for two reasons. First, as illustrated in Fig.~\ref{fig:data_logic_separation}, COBOL separates executable logic in the procedure division from identifier definitions in the data division. When taking only section itself as input for summarization, it misses key properties such as \texttt{PIC} and \texttt{VALUE}; while simply retrieving these definitions is insufficient: their behavioral effects depend on how identifiers participate in comparisons, arithmetic, assignments, and state updates.
Second, a concise summary cannot report every recovered constraint. The model must identify and prioritize constraints that materially affect migration, such as numeric precision, overflow, boundary conditions, and state effects. Effective modernization-oriented summarization therefore requires both reliable constraint recovery and interpretation, and selective preservation of migration-critical constraints.

In this paper, we present \approach, an integrated framework for modernization-oriented COBOL section summarization that addresses both challenges through execution-validated data synthesis and constraint-aware model training. We target COBOL section-level summarization, since this level organizes related paragraphs into a more complete unit of procedural logic, providing a natural granularity for developer-facing documentation~\cite{ibm_cobol_program_format}. To address data scarcity, \approach repurposes the extensive natural-language descriptions and executable tests in general-purpose programming tasks; it uses the former to guide COBOL program generation and the latter to validate and refine the generated programs, thereby enabling execution-validated code-summary supervision without existing COBOL summaries. To preserve migration-critical constraints, \approach transfers the required capabilities to two smaller task-specific models, \emph{model-explain} and \emph{model-summary}. \emph{Model-explain} is fine-tuned on teacher-generated identifier explanations to interpret relevant data definitions and construct expanded context. \emph{Model-summary} is fine-tuned on judge-validated structured rationales to learn which constraints are important for modernization and how to reflect them in the final summary. At inference, the two smaller models run sequentially and locally. They first recover relevant identifier semantics and then generate a constraint-aware summary from the expanded context, without invoking the large teacher LLMs, thereby supporting the data-privacy requirements of enterprise legacy-system modernization.

We evaluate \approach on datasets from both open-source and real-world enterprise COBOL systems (dubbed Stack-120 and Industrial-200). The synthesis stage produces 3,764 execution-validated COBOL code-summary instances for subsequent model training. Built on four base LLMs (with 7B or 8B parameters), \approach improves COBOL summarization performance, with average relative gains of 25.38\%, 53.84\%, and 37.22\% on ROUGE-L, METEOR, and chrF, respectively. Its best configurations, built on these small models, also remain competitive with substantially larger LLMs (with over 284B parameters). In real-world enterprise evaluation, \approach, built on only \modelname{Qwen3-8B}, outperforms the enterprise-deployed \modelname{Qwen3-235B} by 4.35\%, 8.06\%, and 4.21\% in accuracy, completeness, and conciseness, respectively. In ablation experiment, both execution-validated data synthesis and constraint-aware model improves the effectiveness of \approach significantly. These results confirms that constraint-aware reasoning with synthesized data by \approach enables small, locally deployable LLMs for privacy-sensitive COBOL modernization.

The main contributions of this study are as follows.

\begin{itemize}

    \item We present \approach, an integrated framework for modernization-oriented COBOL section summarization that combines execution-validated data synthesis with constraint-aware reasoning. It constructs training supervision without existing COBOL summaries and explicitly recovers and incorporates data constraints that affect section behavior.

    \item We conduct extensive experiments on open-source and real-world enterprise COBOL systems, demonstrating the effectiveness of \approach and its key designs. 

    \item We release replication artifacts to support reproducibility and future research on low-resource legacy languages\footnote{\url{https://github.com/LH01/CoSTAR}}.

\end{itemize}

The remainder of this paper is organized as follows. Section~\ref{sec:related_work} reviews related work. Section~\ref{sec:approach} presents the \approach framework. Section~\ref{sec:evaluation} describes the experimental setup and reports the evaluation results. Section~\ref{sec:threats} discusses threats to validity. Finally, Section~\ref{sec:conclusion} concludes the paper.
\section{Related Work}
\label{sec:related_work}

\subsection{Code Summarization}
\label{sec:related_code_summarization}

Automatic code summarization has evolved from information-retrieval and template-based techniques to neural models, structure-aware approaches, pretrained code models, and, more recently, LLM-based methods~\cite{sun2025source,zhang2024review}. Most data-driven approaches rely on aligned code--summary data, while existing datasets and studies remain concentrated on mainstream languages such as Java and Python~\cite{ahmed2022multilingual,zhang2024review}. Prior work has reduced annotation dependence through self-supervised pretraining, multilingual learning, data augmentation, and LLM-generated summaries for existing code~\cite{niu2022spt,ahmed2022multilingual,shen2024bash,song2024code,su2024distilled}. These approaches, however, still presuppose access to target-language code. For legacy languages such as COBOL, however, commercial confidentiality limits access to real-world code, while sparse fine-grained documentation leaves even fewer aligned code--summary pairs. Consequently, methods that rely on existing target-language code or seed code--summary data for augmentation or pseudo-labeling are difficult to apply directly. \approach addresses this cold-start setting by synthesizing and execution-validating COBOL code--summary data from the natural-language descriptions and executable tests available in general-purpose programming tasks.

Beyond data availability, summarization quality also depends on the program context available to the model. Prior work has modeled AST relations and control flow, augmented prompts with automatically extracted semantic facts, expanded code snippets with variable-related statements, and incorporated calling context~\cite{tang2022ast,shi2023coss,ahmed2024automatic,guo2023snippet,su2025context}. These studies show that the target code alone may be insufficient for complete summarization. However, they primarily capture syntax, control flow, data flow, or call relationships in mainstream languages. \approach instead targets COBOL's separation of data definitions and executable logic by retrieving relevant data declarations for each target section and further explaining the corresponding identifiers in natural language, providing focused cross-division context.

Recent LLM-based work has also explored zero- and few-shot prompting, semantic augmentation, and knowledge distillation for code summarization~\cite{ahmed2022few,sun2025source,su2024distilled,su2025context}. Distillation can train smaller, locally deployable models using summaries generated by larger models, while advanced prompting strategies such as chain-of-thought do not consistently improve summarization across models and languages~\cite{sun2025source,su2024distilled}. \approach goes beyond training on teacher-generated final summaries by using constraint-guided structured rationales under judge-based quality control as intermediate supervision, enabling smaller models to learn how identifier semantics and data constraints shape program behavior.

\subsection{Legacy System Understanding for Modernization}
\label{sec:related_legacy_understanding}

Legacy system modernization spans re-engineering and migration to new languages, architectures, databases, and platforms~\cite{assunccao2025contemporary}. Across these strategies, engineers typically need to recover implemented business functions, component interactions, data and state changes, and business rules that must be preserved. Such knowledge supports feasibility assessment, system decomposition, and target-system design~\cite{ganesan2016survey,ganesan2018formal}. In practice, however, the required expertise is often scarce and existing documentation incomplete~\cite{khadka2014professionals}.

Traditional work supports legacy-system understanding through reverse engineering, program analysis, architecture recovery, business-rule extraction, and redocumentation~\cite{sneed2019cobol,tamburri2018general,moser2021eknows,geist2021leveraging}. These techniques recover artifacts such as architecture views, call and dependency relationships, control- and data-flow representations, data dictionaries, and business rules~\cite{ganesan2018formal,moser2021eknows}. While such artifacts support system-level understanding, engineers maintaining, refactoring, or migrating specific code units also need concise accounts of their behavior, data effects, and governing constraints. Unit-level natural-language summaries therefore complement system-recovery artifacts.

LLMs have recently been applied to legacy-code documentation. Diggs et al. study line-wise comments for MUMPS and mainframe assembly~\cite{diggs2025leveraging}, while XMainframe targets mainframe knowledge and COBOL summarization~\cite{dau2024xmainframe}. 
However, in COBOL, section is the natural functional unit for understanding business logic. A section in the procedure division can organize multiple related paragraphs, forming a more complete yet still localized unit of procedural logic and thus a natural granularity for summarization~\cite{ibm_cobol_program_format}.
To our knowledge, existing legacy-code documentation work has not specifically investigated COBOL section-level summarization, nor how to recover relevant data division constraints and semantics at this granularity, and ensure that these constraints inform the final summary. Such constraint-aware section-level documentation is important for efficient program understanding and behavior-preserving migration of COBOL legacy systems.

\section{Approach}
\label{sec:approach}

\subsection{Overview}
\label{sec:approach_overview}

\begin{figure*}[!t]
    \centering
    \includegraphics[width=0.82\textwidth]{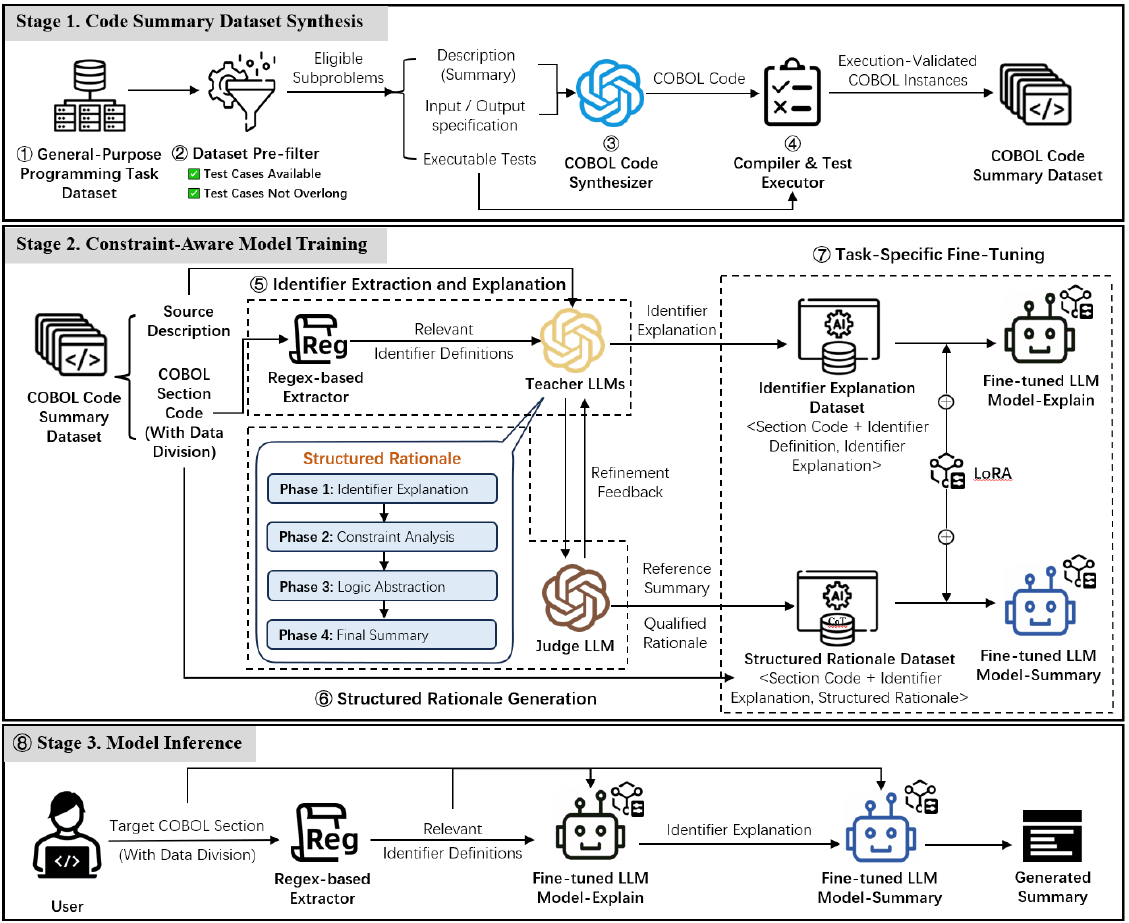}
    \caption{Overview of \approach. Stage~1 synthesizes execution-validated COBOL code--summary data from general-purpose programming tasks. Stage~2 constructs constraint-aware supervision through identifier extraction and explanation and structured rationale generation, and fine-tunes two task-specific models. Stage~3 sequentially applies \emph{model-explain} and \emph{model-summary} to generate the final summary. Teacher and judge LLMs are used only for offline supervision construction and quality control. Circled numbers \circlenum{1}--\circlenum{8} correspond to the core steps described in the text.}
    \label{fig:costar_framework}
\end{figure*}

This section presents \approach, an integrated framework for COBOL section summarization in legacy system modernization. Given a COBOL program containing a data division and a target section in the procedure division, our goal is to generate a concise natural-language summary that captures the section's principal behavior, data effects, and behavior-relevant data constraints. As shown in Fig.~\ref{fig:costar_framework}, \approach comprises three connected stages. Stage 1 repurposes general-purpose programming tasks to synthesize execution-validated COBOL code--summary data, addressing data scarcity. The resulting dataset then supports Stage 2, which connects separated data and logic through relevant data declarations and identifier explanations, and uses constraint-guided structured rationales to train two smaller task-specific models. Stage 3 first explains relevant identifiers in the target section to construct expanded context, and then generates the final constraint-aware summary from this context.

\subsection{Code Summary Dataset Synthesis}
\label{sec:dataset_synthesis}

Stage 1 of \approach constructs COBOL section summarization data from general-purpose programming tasks. It is based on a core observation that a programming task can be represented by a task specification, a program implementation, and executable tests:
\begin{equation}
\mathcal{T}_i =
\langle \mathrm{Spec}_i,\mathrm{Prog}_i,\mathrm{Tests}_i \rangle,
\quad
\mathrm{Spec}_i =
\langle \mathrm{Desc}_i,\mathrm{IO}_i \rangle.
\label{eq:synthesis_triplet}
\end{equation}
Here, $\mathcal{T}_i$ denotes the $i$-th programming task. $\mathrm{Spec}_i$ denotes its task specification, comprising a natural-language description $\mathrm{Desc}_i$ and an input/output specification $\mathrm{IO}_i$; $\mathrm{Prog}_i$ denotes a program implementation; and $\mathrm{Tests}_i$ denotes its executable tests. The specification and tests jointly constrain the expected program behavior, providing both guidance for program synthesis and executable evidence for validating the generated implementation.

\approach instantiates this relation by providing $\mathrm{Spec}_i$ to a Code LLM to synthesize a COBOL implementation $\mathrm{Prog}_i$, while withholding $\mathrm{Tests}_i$ for subsequent execution validation. This repurposes the natural-language descriptions and executable tests already available in general-purpose programming tasks to construct execution-validated COBOL code--summary supervision. As shown in Stage 1 of Fig.~\ref{fig:costar_framework}, the process comprises four core elements: \circlenum{1} a general-purpose programming task dataset, \circlenum{2} a dataset pre-filter, \circlenum{3} a COBOL code synthesizer, and \circlenum{4} a compiler \& test executor. We describe them below in the same order.

\subsubsection{General-Purpose Programming Task Dataset}
\label{sec:general_purpose_programming_dataset}

The synthesis stage does not assume a fixed source-task granularity. Since this study uses COBOL sections as the function-level summarization unit, we instantiate the process with function-level programming tasks. Specifically, we use CodeFlowBench~\cite{wang2026codeflowbench}, whose snapshot in our study contains 11,592 function-level subproblems, providing sufficient synthesis scale and a natural granularity match with individual COBOL sections. Each subproblem provides a natural-language task description, an input/output specification, and executable tests, supplying $\mathrm{Spec}_i$ and $\mathrm{Tests}_i$ in Eq.~\ref{eq:synthesis_triplet}. The synthesis process 
can be applied to other programming-task resources that provide comparable task specifications and executable tests.

\subsubsection{Dataset Pre-filter}
\label{sec:dataset_pre_filter}

Before synthesis, the dataset pre-filter removes subproblems that either lack executable tests or contain excessively long test content. The former cannot support behavioral validation, while the latter incur excessive processing overhead. We define the eligible subset as
\begin{equation}
\mathcal{D}_{\mathrm{elig}}
=
\{q_i\in\mathcal{D}_{\mathrm{raw}}
\mid
\mathrm{Tests}_i\neq\emptyset,\,
\ell_{\mathrm{tok}}(\mathrm{Tests}_i)\le\tau\}.
\label{eq:eligible_filter}
\end{equation}

\noindent Here, $q_i$ denotes the $i$-th source subproblem; $\mathcal{D}_{\mathrm{raw}}$ and $\mathcal{D}_{\mathrm{elig}}$ denote the original and eligible subproblem sets, respectively; $\mathrm{Tests}_i$ denotes the executable tests associated with $q_i$; $\ell_{\mathrm{tok}}(\mathrm{Tests}_i)$ denotes their token length; and $\tau$ is the maximum allowed length.

Based on the empirical distribution of test-content lengths, we set $\tau$ to 5,000 tokens to remove a small number of unusually long cases while retaining the vast majority of subproblems with executable tests. Among the 11,592 source subproblems, 528 are removed for lacking executable tests and another 140 for exceeding the threshold, leaving 10,924 eligible subproblems.

\begin{table}[t]
\centering
\caption{Statistics of code summary dataset synthesis.}
\label{tab:dataset_synthesis_statistics}
\footnotesize
\renewcommand{\arraystretch}{1.15}
\setlength{\tabcolsep}{6pt}
\renewcommand{\tabularxcolumn}[1]{m{#1}}

\begin{tabularx}{0.85\linewidth}{
@{\hspace{10pt}}
>{\centering\arraybackslash}X
>{\centering\arraybackslash}c
@{\hspace{10pt}}
}
\toprule
\textbf{Item} & \textbf{Count} \\
\midrule
Original subproblems & 11,592 \\
Removed without executable tests & 528 \\
Removed with overlong test content & 140 \\
Eligible subproblems & 10,924 \\
Execution-validated COBOL instances & 3,764 \\
\bottomrule
\end{tabularx}
\end{table}

After pre-filtering, the task specification of each eligible subproblem is passed to the COBOL code synthesizer, while its executable tests are reserved for subsequent validation.

\subsubsection{COBOL Code Synthesizer}
\label{sec:cobol_code_synthesizer}

\begin{figure}[t]
    \centering
    \includegraphics[width=0.84\columnwidth]{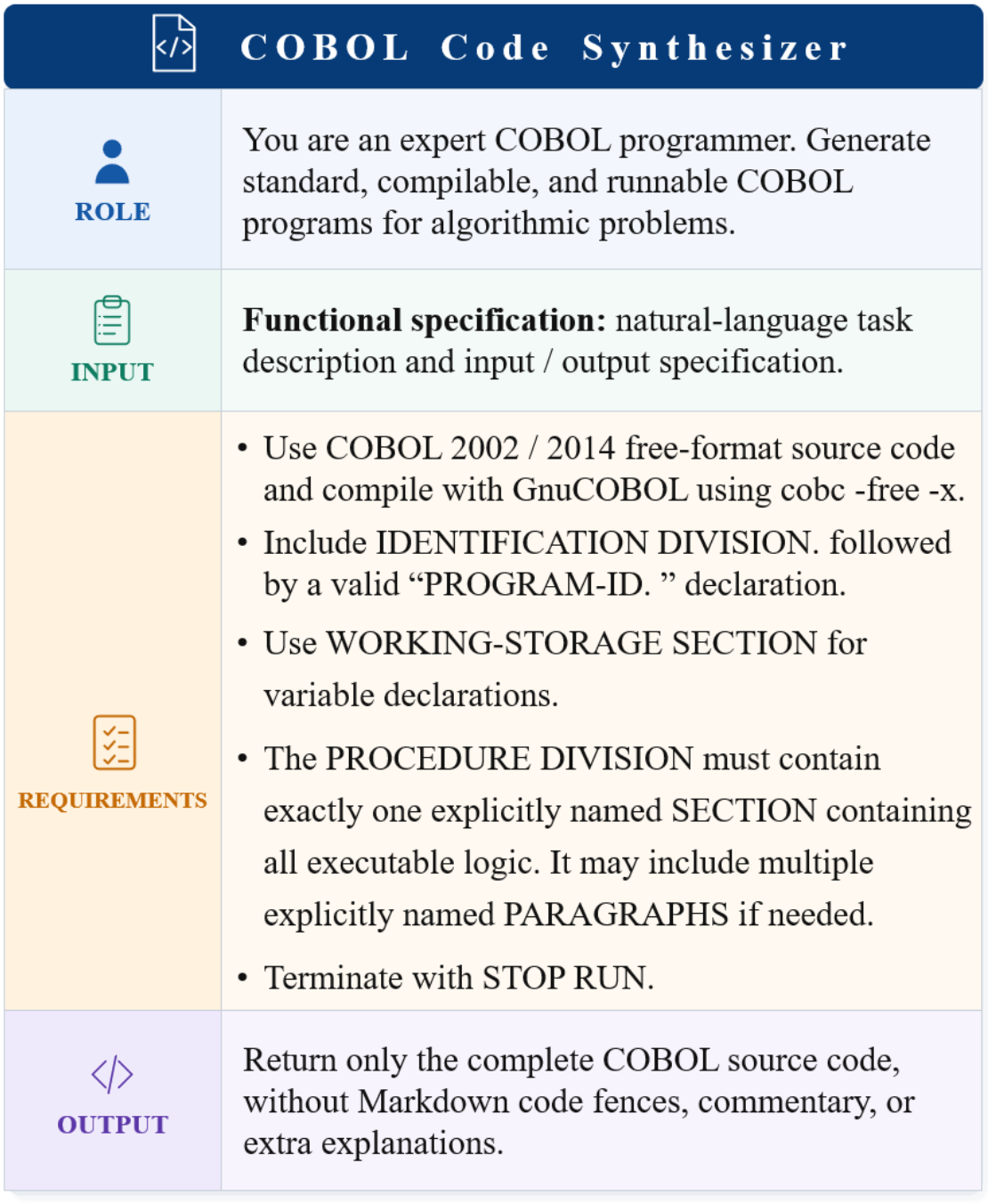}
    \caption{Condensed prompt used by the COBOL code synthesizer. The complete prompt is available in our replication repository.}
    \label{fig:cobol_synthesis_prompt}
\end{figure}

For each eligible subproblem, the COBOL code synthesizer instantiates a predefined prompt with its natural-language task description and input/output specification and submits the prompt to a Code LLM. As shown in Fig.~\ref{fig:cobol_synthesis_prompt}, the prompt specifies the COBOL source format, program structure, input/output conventions, and section organization required for subsequent compilation, execution validation, and section-level summarization. The executable tests are deliberately withheld from the Code LLM and used only by the subsequent compiler \& test executor.

The synthesizer generates a complete COBOL program rather than an isolated section. This provides both the complete structure required for compilation and execution and the data division declarations needed for subsequent constraint-aware summarization. Since a task specification may admit multiple correct implementations, each generated program is treated as a candidate until it passes execution validation. The source task description is retained as the paired summary for the target section once the program is validated.

\subsubsection{Compiler \& Test Executor}
\label{sec:compiler_test_executor}

The compiler \& test executor validates each candidate through compilation and test execution. Let $\mathrm{Compile}_i$ indicate that candidate $\mathrm{Prog}_i$ compiles successfully and $\mathrm{PassAll}_i$ that it passes every executable test in $\mathrm{Tests}_i$. We define the acceptance indicator as
\begin{equation}
\operatorname{Accept}_i =
\begin{cases}
1, & \mathrm{Compile}_i \land \mathrm{PassAll}_i,\\
0, & \text{otherwise}.
\end{cases}
\label{eq:validation_rule}
\end{equation}

We compile each candidate with GnuCOBOL 3.2 in free source format using \texttt{cobc -free -x}. Its open-source command-line compiler supports the required free-format COBOL source and can be readily integrated into our automated execution-validation pipeline.

Compilation and testing provide complementary validation signals. Compilation verifies that the generated program is accepted by the target compiler, while testing checks whether its observable behavior satisfies the source task requirements. Passing a finite test suite does not prove complete semantic correctness, but provides executable evidence that each accepted program satisfies all tested behaviors.

Each accepted program yields one execution-validated COBOL section instance, with the source task description serving as its paired summary and the target section, data division, and other information required for subsequent data construction retained. As shown in Table~\ref{tab:dataset_synthesis_statistics}, 3,764 of the 10,924 eligible subproblems yield programs that compile successfully and pass all tests. These instances form the COBOL code summary dataset that serves as the data foundation for constraint-aware model training in Stage 2 of \approach.

\subsection{Constraint-Aware Model Training}
\label{sec:costar}
Stages~2 of \approach address the second challenge, the separation of data and logic. Rather than merely linking the data and procedure divisions, \approach progressively recovers relevant data definitions, explains their semantics, reasons about how the resulting constraints affect section behavior, and transfers this process to two smaller task-specific models. As shown in Fig.~\ref{fig:costar_framework}, Stage~2 comprises three dependent steps. Step~\circlenum{5} recovers and explains relevant data definitions to construct expanded context; Step~\circlenum{6} builds constraint-guided reasoning supervision from this context; and Step~\circlenum{7} uses the resulting data to fine-tune two smaller task-specific models (i.e., \emph{model-explain} and \emph{model-summary}). Strong teacher LLMs are used in Steps~\circlenum{5} and~\circlenum{6} to construct supervision, and the two teacher roles need not use the same underlying model; a judge LLM performs quality control. These large models are used only offline.

\subsubsection{Identifier Extraction and Explanation}
\label{sec:identifier_extraction_explanation}

Constraint-aware summarization first requires recovering the data definitions on which the target section actually depends. For each execution-validated COBOL section instance produced in Stage~1, the regex-based extractor identifies data identifiers referenced by the target section and retrieves their corresponding declarations from the data division, while retaining necessary parent group items and key defining information such as \texttt{PIC} and \texttt{VALUE}. \texttt{PIC} describes the category and format of a data item, including properties such as field width and implied decimal positions, whereas \texttt{VALUE} specifies its initial value. Such definitions can affect comparisons, assignments, boundary handling, and other program behavior, making them important context for understanding the section. We refer to the retrieved declarations as \emph{relevant identifier definitions}, which preserve pertinent data constraints while excluding unrelated data division declarations.

However, raw COBOL declarations remain compact and symbolic, making their data semantics and potential risks difficult for the smaller, locally deployable models targeted in this work to interpret reliably. To provide these models with explicit semantic supervision, a stronger teacher LLM generates a natural-language identifier explanation for each retrieved identifier from its name and raw COBOL definition, describing its physical semantics, constraint bounds, and potential risk profile. The resulting explanations are then combined with the target section and relevant identifier definitions to form the expanded context for subsequent constraint-aware reasoning.

The generated results also form the identifier explanation dataset. Each sample maps an identifier and its raw definition to the corresponding identifier explanation. Step~\circlenum{7} later uses this dataset to train \emph{model-explain}, enabling the same identifier-level semantics and constraints to be recovered individually without invoking a teacher LLM during inference.

\subsubsection{Structured Rationale Generation}
\label{sec:structured_rationale_generation}

The identifier explanation supplies the target section with relevant data semantics, but providing this information alone does not ensure that a summarization model will correctly reason about its behavioral implications or preserve important constraints in the final summary. We therefore construct structured rationale supervision that explicitly organizes identifier understanding, constraint analysis, logic abstraction, and final summarization into a progressive reasoning process.

The Description retained in the COBOL code summary dataset originates from the source general-purpose programming task and specifies the intended program behavior. Because it exists before the concrete COBOL implementation is synthesized, it cannot capture the data definitions and constraints introduced by that implementation and its data division. \approach therefore uses the Description only as an auxiliary behavioral specification during offline structured rationale construction rather than directly treating it as the final constraint-aware reference summary.

A teacher LLM uses the expanded context together with this behavioral specification to generate a structured rationale organized into four ordered phases:

\begin{enumerate}
    \item \textbf{Identifier Explanation} identifies the roles and behavior-relevant properties of the data items used by the section.

    \item \textbf{Constraint Analysis} examines COBOL-specific constraints such as field formats, numeric precision, signedness, initialization values, hierarchy, and shared-state updates.

    \item \textbf{Logic Abstraction} summarizes the section's control flow, data flow, branch conditions, iterations, external interactions, and state changes.

    \item \textbf{Final Summary} integrates the preceding three phases to generate a concise summary of the section's principal behavior, data effects, and migration-relevant constraints.
\end{enumerate}

This organization grounds the final summary in explicitly recovered identifier semantics, data constraints, and section behavior rather than only surface patterns in the local code. In particular, the fourth phase, \emph{Final Summary}, generates the summary from the preceding three phases so that it can incorporate constraints introduced by the concrete COBOL implementation rather than directly reuse the source Description.

Automatically generated structured rationales may still contain interpretations inconsistent with the code context, incoherent reasoning, or omissions of important constraints. To prevent such errors from entering the fine-tuning data, a judge LLM evaluates each candidate and returns refinement feedback when the required quality criteria are not satisfied.

Given the target section, relevant identifier definitions, identifier explanation, and candidate structured rationale, including its final summary, the judge LLM evaluates the result according to the three criteria summarized in Table~\ref{tab:judge_criteria}.

\begin{table}[t]
\centering
\caption{Judge criteria and acceptance thresholds.}
\label{tab:judge_criteria}
\footnotesize
\renewcommand{\arraystretch}{1.05}
\setlength{\tabcolsep}{2pt}
\renewcommand{\tabularxcolumn}[1]{m{#1}}

\begin{tabularx}{\linewidth}{
@{}
>{\centering\arraybackslash}m{0.23\linewidth}
>{\raggedright\arraybackslash}X
>{\centering\arraybackslash}m{0.12\linewidth}
@{}
}
\toprule
\textbf{Criterion} &
\multicolumn{1}{c}{\textbf{Evaluation Focus}} &
\textbf{Threshold} \\
\midrule

Context Adherence &
Whether identifier explanations are grounded in the provided code and declarations without unsupported properties or omissions. &
$\geq 4$ \\

\specialrule{0.3pt}{1.2pt}{1.2pt}

Logical Coherence &
Whether constraint analysis and logic abstraction form a consistent reasoning process. &
$\geq 4$ \\

\specialrule{0.3pt}{1.2pt}{1.2pt}

Constraint Coverage &
Whether the final summary preserves the migration-relevant constraints and state effects identified in the rationale. &
$= 5$ \\

\bottomrule
\end{tabularx}
\end{table}

Each criterion is scored on a five-point scale. A candidate is accepted only when Context Adherence and Logical Coherence both score at least 4 and Constraint Coverage receives a score of 5. The first two thresholds allow only minor deficiencies that do not alter the core interpretation, whereas Constraint Coverage uses the strictest threshold because omitting an identified migration-relevant constraint or state effect would directly weaken subsequent summarization supervision.

A candidate that fails any threshold is returned to the teacher together with judge-generated refinement feedback. The teacher revises the structured rationale according to this feedback and resubmits it for evaluation, forming a generate--judge--refine loop summarized in Algorithm~\ref{alg:rationale_refinement}.

\begin{algorithm}[t]
\caption{Teacher--Judge Structured Rationale Refinement}
\label{alg:rationale_refinement}
\small
\begin{algorithmic}[1]
\Require Instance context $x$; Teacher $T$; Judge $J$
\Require Maximum generation rounds $G$; maximum refinements $R$
\Ensure Qualified structured rationale $r$, or $\emptyset$

\For{$g \gets 1$ to $G$}
    \State $r \gets T.\Call{Generate}{x}$
    \For{$k \gets 0$ to $R$}
        \State $(q,f) \gets J.\Call{Evaluate}{x,r}$
        \If{$q$ satisfies all acceptance thresholds}
            \State \Return $r$
        \EndIf
        \If{$k < R$}
            \State $r \gets T.\Call{Refine}{x,r,f}$
        \EndIf
    \EndFor
\EndFor
\State \Return $\emptyset$
\end{algorithmic}
\end{algorithm}

In our implementation, each generation round allows at most three teacher--judge refinements, and each instance allows at most five independent generation rounds. Local refinement revises an existing candidate according to judge feedback; if the candidate remains unqualified after reaching the refinement limit, a new candidate is generated independently from the original inputs. If a structured-rationale supervision sample remains unqualified across all generation rounds, it is excluded from the structured rationale dataset.

Once a candidate passes quality control, its complete four-phase output is retained as the qualified rationale, with the Phase~4 final summary serving as the reference summary. Thus, the qualified rationale and reference summary entering the structured rationale dataset in Fig.~\ref{fig:costar_framework} originate from the same quality-controlled output rather than from the source Description.

\begin{figure*}[t]
    \centering
    \includegraphics[width=0.82\textwidth]{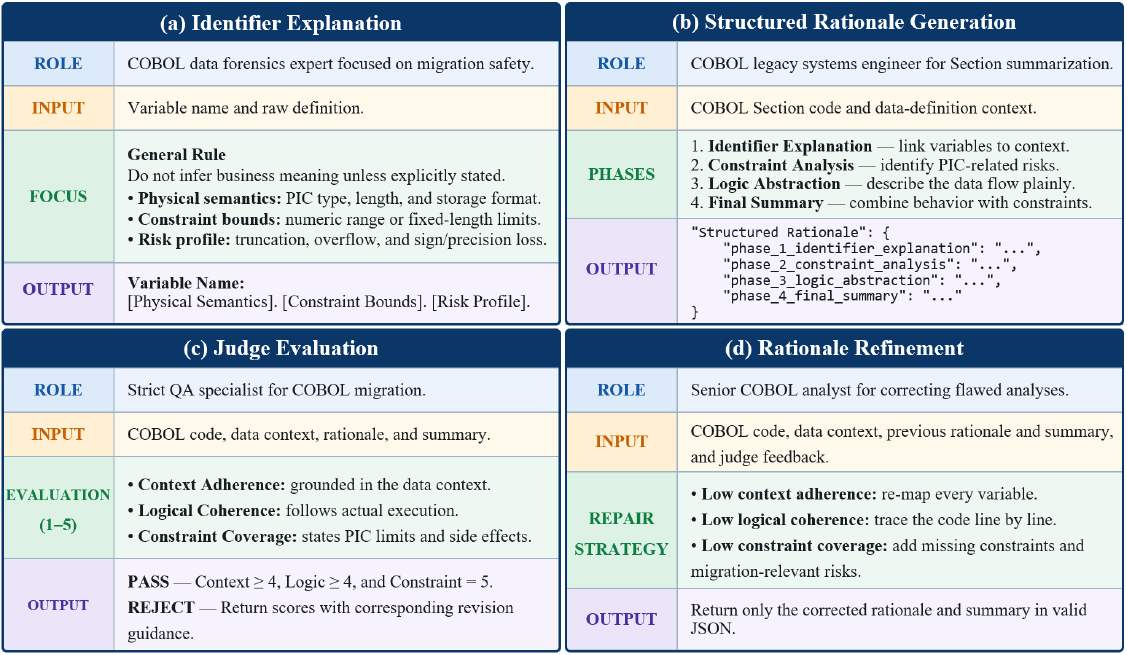}
    \caption{Condensed prompt templates used for offline supervision construction in \approach: (a) Identifier Explanation, (b) Structured Rationale Generation, (c) Judge Evaluation, and (d) Rationale Refinement. Complete prompts are provided in our replication repository.}
    \label{fig:costar_prompts}
\end{figure*}

The key prompt instructions for identifier explanation, structured rationale generation, judge evaluation, and rationale refinement are summarized in Fig.~\ref{fig:costar_prompts}. The figure retains only the instructions defining the inputs, core tasks, and output structures; the complete prompt templates are provided in our replication repository.

\subsubsection{Task-Specific Fine-Tuning}
\label{sec:task_specific_fine_tuning}

Steps~\circlenum{5} and~\circlenum{6} produce two connected forms of supervision. The identifier explanation dataset teaches a model to recover identifier semantics from each identifier and its corresponding definition within the target section context, while the structured rationale dataset teaches a second model to perform constraint-aware reasoning and summarization from the expanded context.

We perform supervised fine-tuning with low-rank adaptation (LoRA) on the two datasets, producing two task-specific models. \emph{model-explain} learns to generate the identifier explanation, whereas \emph{model-summary} learns to generate the four-phase structured rationale from the expanded context, with Phase~4 providing the final constraint-aware summary. Through this generated supervision, task-specific COBOL understanding and constraint reasoning from the stronger teacher LLMs are transferred to the two smaller deployable models.

This task decomposition separates identifier-semantic recovery from constraint-aware summarization rather than requiring a single smaller model to learn the entire mapping from raw COBOL code to a constraint-aware summary. Neither the source Description nor judge-generated refinement feedback is available during inference; the teacher and judge LLMs are used only for offline supervision construction and quality control.

\subsection{Model Inference}
\label{sec:constraint_aware_inference}

Stage~3 performs Step~\circlenum{8}, sequentially applying the two task-specific models trained in Stage~2 to previously unseen COBOL sections. As shown in Fig.~\ref{fig:costar_framework}, the user provides only the target section and its data division; inference requires neither the Description from the source general-purpose programming task nor the teacher and judge LLMs.

The same regex-based extractor first retrieves the relevant identifier definitions for identifiers referenced by the target section from the data division. \emph{model-explain} then processes each retrieved identifier individually, taking its name and raw definition as input to generate the corresponding identifier explanation. The explanations of all relevant identifiers are then aggregated with the target section and relevant identifier definitions to form the expanded context. \emph{model-summary} generates a structured rationale from this context following the four-phase organization learned during fine-tuning, and its Phase~4 final summary is returned to the user as the generated summary.

Consequently, deployment requires only the cross-division information relevant to the current section and the sequential application of two smaller task-specific models, without invoking the large teacher or judge LLMs. This design explicitly incorporates data constraints that affect section behavior while confining actual inference to locally deployable smaller models, making it suitable for privacy-sensitive enterprise COBOL environments.
\section{Evaluation}
\label{sec:evaluation}

To comprehensively evaluate \approach, we first examine its overall effectiveness, then investigate the contributions of synthesized data, expanded context, and structured rationale supervision, and finally assess its applicability to real-world enterprise COBOL modernization. We investigate the following five research questions.

\begin{itemize}

    \item \textbf{RQ1. Overall Effectiveness:}
    How does \approach perform compared with its corresponding base LLMs and large-scale LLM baselines?

    \item \textbf{RQ2. Synthesized Data Effectiveness:}
    How does synthesized training-data scale affect COBOL section summarization performance?

    \item \textbf{RQ3. Expanded Context:}
    How does the expanded context used during training and inference affect the performance of \approach?

    \item \textbf{RQ4. Structured Rationale Supervision:}
    How does constraint-guided structured rationale supervision affect the performance of \approach compared with summary-only supervised fine-tuning (hereafter Summary-only SFT)?

    \item \textbf{RQ5. Industrial Applicability:}
    How does \approach perform in real-world enterprise COBOL modernization scenarios?

\end{itemize}

\subsection{Experimental Setup}
\label{sec:experimental_setup}

\subsubsection{Datasets}

We use two complementary evaluation datasets covering open-source and real-world enterprise COBOL systems.

\noindent\textbf{Stack-120.}
Stack-120 is our public evaluation dataset for COBOL section-level code summarization. We construct it by manually selecting 120 sections from open-source COBOL programs in The Stack together with their relevant data division declarations.

For each section, one engineer from our industrial partner writes an initial reference summary following a unified annotation guideline, while the other two independently review it against the target section and its relevant data division declarations. Any disagreements are resolved through discussion among all three engineers until consensus is reached. Each engineer has more than ten years of COBOL development experience. The resulting references describe the principal section behavior and preserve modernization-relevant information, including important data items, state effects, data constraints, boundary behavior, and potential migration risks. Stack-120 is used for the automatic evaluation in RQ1--RQ4.

\noindent\textbf{Industrial-200.}
Industrial-200 contains 200 business-critical sections selected by our industrial partner from real-world enterprise COBOL systems undergoing modernization. These sections implement core business logic targeted by the partner's modernization efforts and are used exclusively for the human evaluation in RQ5. Because the source systems contain confidential business logic, Industrial-200 is not publicly released. All model inference and evaluation are conducted within the partner's local environment, and only anonymized ratings are returned for analysis.

For each evaluation instance, we extract the target section and its relevant data division declarations, retaining necessary parent group items and related structural information when available. Unavailable copybook definitions are not inferred. The same preprocessing procedure is applied across comparable models and experimental settings.

\subsubsection{Evaluation Metrics}

Existing evaluation metrics for code summarization can be broadly categorized into textual similarity metrics, semantic similarity metrics, and human evaluation metrics~\cite{wu2025can}. To evaluate \approach, we adopt metrics from all three categories.

\noindent\textbf{Textual Similarity Metrics.}
Following prior studies~\cite{hu2022correlating,haque2022semantic,stapleton2020human,roy2021reassessing}, we report ROUGE-L, METEOR, and chrF. ROUGE-L measures sequence-level overlap based on the longest common subsequence. METEOR measures unigram alignment while considering matching order and fragmentation. chrF measures character $n$-gram similarity and is particularly suitable for summaries containing identifiers, abbreviations, and numeric expressions. Although BLEU is frequently used in code summarization, prior studies show that it can be unstable and weakly correlated with human judgments~\cite{hu2022correlating,haque2022semantic,stapleton2020human}. We therefore do not use BLEU as a primary metric.

\noindent\textbf{Semantic Similarity Metrics.}
Following prior studies~\cite{hu2022correlating,haque2022semantic}, we use BERTScore and SentenceBERT similarity. BERTScore evaluates token-level semantic correspondence using contextual embeddings, whereas SentenceBERT measures cosine similarity between sentence-level representations of the generated and reference summaries.

All automatic scores are reported as percentages, with higher values indicating greater similarity to the reference summary. We compute ROUGE-L using \texttt{rouge-score}, METEOR using \texttt{NLTK}, chrF using \texttt{SacreBLEU}, BERTScore using the \texttt{bert-score} library with the pretrained \texttt{RoBERTa-large} model, and SentenceBERT similarity using \texttt{sentence-transformers} with the pretrained \texttt{stsb-roberta-large} model.

\noindent\textbf{Human Evaluation Metrics.}
Automatic metrics cannot fully determine whether a generated summary preserves COBOL-specific constraints, boundary conditions, state effects, and migration-relevant risks~\cite{hu2022correlating,roy2021reassessing}. We therefore conduct human evaluation on Industrial-200, with the detailed protocol presented in RQ5.

\subsubsection{Baselines}

To assess the performance gap between smaller \approach models and large-scale general-purpose LLMs, we include five large-scale LLM baselines: the open-weight \modelname{DeepSeek-V3.2} and \modelname{DeepSeek-V4-Flash}, and the closed-source \modelname{Gemini-2.5-Flash-Lite}, \modelname{Gemini-3.1-Flash-Lite}, and \modelname{Claude-Haiku-4.5}. These models receive no task-specific fine-tuning and use the same task instructions, input information, and output requirements as \approach whenever supported. The base LLMs and controlled comparison settings used for individual RQs are introduced within the corresponding RQs.

\subsubsection{Implementation Details}

The offline data-construction pipeline of \approach does not depend on specific LLM choices. In our implementation, we instantiate its code-generation, supervision-generation, and quality-control roles with \modelname{DeepSeek-V3.2}, \modelname{Qwen3-Coder-480B-A35B-Instruct}, and \modelname{DeepSeek-R1}, respectively, and keep these implementation choices fixed across all experiments to avoid introducing additional model-selection variation. Specifically, \modelname{DeepSeek-V3.2} generates COBOL programs in Stage~1, \modelname{Qwen3-Coder-480B-A35B-Instruct} generates identifier explanations and structured rationales in Stage~2, and \modelname{DeepSeek-R1} performs quality evaluation. These models are used only for offline data construction, while inference uses the fine-tuned task-specific models. Generated COBOL programs are compiled with GnuCOBOL~3.2 in free source format using \texttt{cobc -free -x <source\_file> -o <program\_name>}, and only programs that compile successfully and pass all associated tests are retained.

All fine-tuning experiments are conducted on a server running Ubuntu 22.04.4 LTS with one NVIDIA RTX A6000 GPU with 48\,GB of memory, 128\,GB of RAM, and an Intel Core i9-13900K CPU. We implement the training pipeline using LLaMA-Factory 0.9.5.dev0, Python 3.11.14, and PyTorch 2.10.0.

For each \approach instantiation, \emph{model-explain} and \emph{model-summary} are independently fine-tuned from the same base LLM to control backbone variation, although the framework itself does not require them to share the same backbone. Both models use LoRA with rank 8, scaling factor 16, dropout 0, and a cutoff length of 4,096. We train for three epochs with a learning rate of $5 \times 10^{-5}$, a per-device batch size of 2, eight gradient accumulation steps, a maximum gradient norm of 1.0, and BF16 precision. We reserve 15\% of the training data for validation and use the final checkpoint after the third epoch.

Across all controlled experiments, we keep the training and inference settings unchanged except for the factor explicitly examined by the corresponding RQ, thereby isolating its effect on summarization performance.

During inference, all models use the same preprocessing and post-processing procedures. Local models use nucleus sampling with a temperature of 0.3 and a top-$p$ value of 0.95. Large-scale LLM baselines are accessed through public APIs and use the same task prompt and decoding settings whenever supported. For \approach, all automatic and human evaluations use only the Phase~4 final summary; the intermediate structured rationale is excluded from metric computation.

\subsection{RQ1: Overall Effectiveness}
\label{sec:rq1}

\noindent\textbf{Motivation.}
As an integrated framework, \approach should first demonstrate its overall effectiveness across different base LLMs. We therefore examine whether \approach consistently improves different base LLMs and whether its variants built on small-scale LLMs (e.g., 7B or 8B models) can compete with substantially larger general-purpose LLMs.

\noindent\textbf{Methodology.}
Considering model popularity and broad adoption, representativeness across model types, and feasibility for local deployment, we select four well-known open-weight 7B or 8B base LLMs from different model families: the code-oriented \modelname{CodeGemma-1.1-7B-Instruct}, the reasoning-oriented \modelname{DeepSeek-R1-Distill-Llama-8B}, and the general instruction models \modelname{Llama-3.1-8B-Instruct} and \modelname{Qwen3-8B}. This selection allows us to examine the applicability of \approach across different model types and families while controlling the model scale. We first compare each \approach variant with its corresponding original base LLM to isolate the gains introduced by the complete framework. We then compare the \approach variants with the five large-scale LLM baselines described in Section~\ref{sec:experimental_setup}. All models are evaluated on Stack-120 using the five automatic metrics.

\begin{table*}[t]
\centering
\begin{threeparttable}

\caption{Overall performance of \approach{} compared with base LLMs and large-scale LLM baselines on Stack-120.}
\label{tab:overall_performance}

\small
\setlength{\tabcolsep}{2pt}
\renewcommand{\arraystretch}{1.12}
\sisetup{
    table-number-alignment = center,
    reset-text-series = false,
    text-series-to-math = true,
    reset-text-family = false,
    text-family-to-math = true
}

\begin{tabular*}{\textwidth}{
    @{\extracolsep{\fill}}
    >{\centering\arraybackslash}m{2.45cm}
    >{\centering\arraybackslash}m{4.80cm}
    >{\centering\arraybackslash}m{0.70cm}
    S[table-format=2.3, table-column-width=1.35cm]
    S[table-format=2.3, table-column-width=1.35cm]
    S[table-format=2.3, table-column-width=1.20cm]
    S[table-format=2.3, table-column-width=1.55cm]
    S[table-format=2.3, table-column-width=2.20cm]
    @{}
}
\toprule
\textbf{Category}
& \textbf{Model}
& \textbf{Size}
& {\textbf{ROUGE-L}}
& {\textbf{METEOR}}
& {\textbf{chrF}}
& {\textbf{BERTScore}}
& {\textbf{SentenceBERT}} \\
\midrule

\multirow{4}{*}{Base LLM}
& CodeGemma-1.1-7B-Instruct
& 7B
& 16.535
& 11.624
& 17.352
& 83.684
& 55.411 \\

& DeepSeek-R1-Distill-Llama-8B
& 8B
& 13.040
& 10.411
& 17.704
& 82.700
& 52.322 \\

& Llama-3.1-8B-Instruct
& 8B
& 22.452
& 18.204
& 25.305
& 84.932
& 61.740 \\

& Qwen3-8B
& 8B
& 23.306
& 20.605
& 28.372
& 85.694
& 64.411 \\

\midrule

\multirow{4}{*}{\approach}
& CodeGemma-1.1-7B-Instruct
& 7B
& 19.550
& 21.555
& 29.368
& 84.354
& 61.136 \\

& DeepSeek-R1-Distill-Llama-8B
& 8B
& 22.048
& 20.102
& 26.710
& 85.029
& 60.464 \\

& Llama-3.1-8B-Instruct
& 8B
& 24.990
& 23.670
& 30.324
& 85.807
& 64.982 \\

& Qwen3-8B
& 8B
& 23.984
& 22.012
& 30.901
& 86.023
& 67.881 \\

\midrule

\multirow{5}{*}{Large-scale LLM}
& DeepSeek-V3.2
& 671B
& 25.865
& 20.060
& 29.709
& 84.910
& 67.468 \\

& DeepSeek-V4-Flash
& 284B
& 28.237
& 23.528
& 32.406
& 85.610
& 70.638 \\

& Gemini-2.5-Flash-Lite
& N/A
& 25.856
& 20.908
& 28.495
& 85.753
& 67.665 \\

& Gemini-3.1-Flash-Lite
& N/A
& 21.065
& 18.921
& 32.355
& 84.659
& 66.367 \\

& Claude-Haiku-4.5
& N/A
& 24.214
& 19.469
& 31.222
& 84.841
& 67.596 \\

\bottomrule
\end{tabular*}

\vspace{0.6ex}

\begin{tablenotes}[flushleft]
\footnotesize
\item[] \textit{Note.} N/A indicates that the parameter size of the closed-source models is not publicly disclosed.
\end{tablenotes}

\end{threeparttable}
\end{table*}

\begin{figure}[!t]
\centering
\includegraphics[
    width=\columnwidth,
    keepaspectratio
]{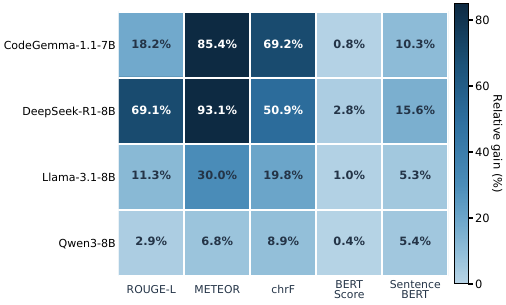}
\caption{Relative gains of \approach over the corresponding base LLMs on Stack-120. Each cell reports the relative improvement obtained by applying \approach to the same underlying base LLM. DeepSeek-R1-8B abbreviates DeepSeek-R1-Distill-Llama-8B.}
\label{fig:rq1_base_gain}
\end{figure}

\begin{figure}[!t]
\centering
\includegraphics[
    width=0.7\linewidth,
    keepaspectratio
]{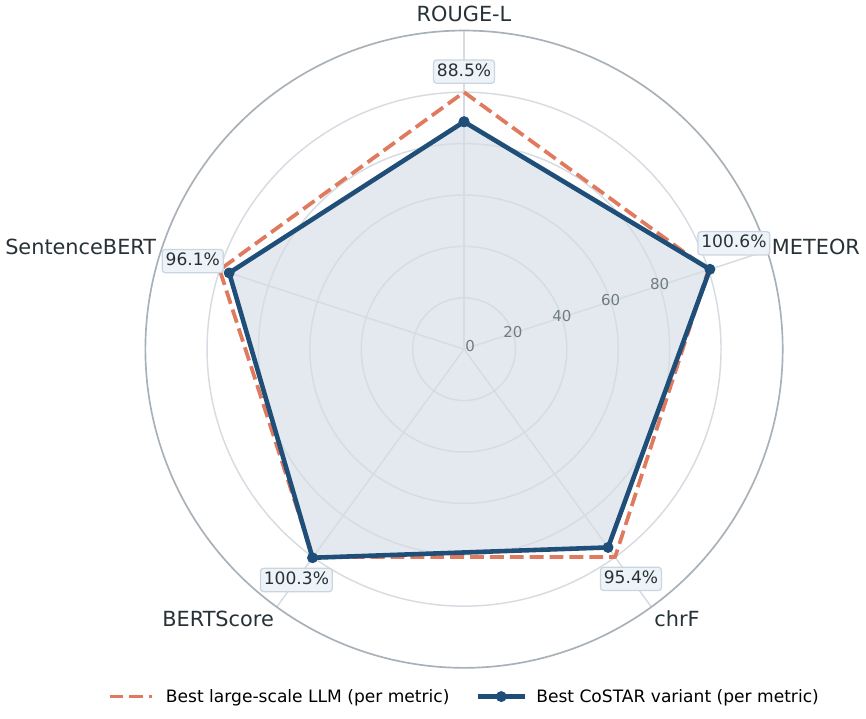}
\caption{Per-metric comparison between the best \approach and the best large-scale LLM baseline on Stack-120. Each metric is normalized to its best large-scale LLM score (set to 100\%); values above 100\% indicate better \approach performance. The top-performing model may vary across metrics.}
\label{fig:rq1_large_scale}
\end{figure}

\noindent\textbf{Results.}
Table~\ref{tab:overall_performance} reports the exact performance of \approach, its corresponding base LLMs, and the large-scale LLM baselines on Stack-120. Fig.~\ref{fig:rq1_base_gain} visualizes the improvements obtained over the corresponding base LLMs, while Fig.~\ref{fig:rq1_large_scale} compares the per-metric best \approach results with the best large-scale LLM baselines.

\noindent\textit{Comparison with Base LLMs.}
As shown in Table~\ref{tab:overall_performance} and Fig.~\ref{fig:rq1_base_gain}, \approach improves every base LLM across all five automatic metrics. Averaged across the four model families, the relative gains are 25.38\% on ROUGE-L, 53.84\% on METEOR, 37.22\% on chrF, 1.26\% on BERTScore, and 9.13\% on SentenceBERT. The larger improvements on the textual similarity metrics indicate closer lexical and structural alignment with the human-written references. The consistent gains on BERTScore and SentenceBERT further show that the improvements are not limited to surface-level overlap but also extend to semantic similarity.

The improvements are observed across model families with different initial performance levels. \modelname{DeepSeek-R1-Distill-Llama-8B} obtains the largest overall gains, including improvements of 69.08\%, 93.08\%, and 50.87\% on ROUGE-L, METEOR, and chrF, respectively. \modelname{CodeGemma-1.1-7B-Instruct} also benefits substantially, with gains of 85.44\% on METEOR and 69.25\% on chrF. Although \modelname{Qwen3-8B} is the strongest original base LLM across all five metrics and consequently exhibits smaller relative gains, \approach still improves every metric. These results demonstrate that the effectiveness of \approach is not confined to a particular model family or to base LLMs with weak initial performance.

\noindent\textit{Comparison with Large-Scale LLMs.}
Despite being instantiated with only 7B or 8B base LLMs, the \approach variants remain competitive with the large-scale LLM baselines. As shown in Fig.~\ref{fig:rq1_large_scale}, the per-metric best \approach results reach 88.50\%, 100.60\%, 95.36\%, 100.31\%, and 96.10\% of the corresponding best large-scale LLM results on ROUGE-L, METEOR, chrF, BERTScore, and SentenceBERT, respectively. In particular, \approach{}-\modelname{Llama-3.1-8B-Instruct} achieves the highest METEOR among all compared models, exceeding the best large-scale LLM baseline by 0.142 points. Similarly, \approach{}-\modelname{Qwen3-8B} achieves the highest overall BERTScore, exceeding the best large-scale LLM baseline by 0.270 points.

This competitiveness is not merely a consequence of selecting a different \approach variant for each metric. A single variant, \approach{}-\modelname{Llama-3.1-8B-Instruct}, outperforms every large-scale LLM baseline on at least two metrics. It exceeds \modelname{DeepSeek-V4-Flash} on METEOR and BERTScore and outperforms each of the other four large-scale baselines on at least three of the five metrics. Although the large-scale LLMs retain the best ROUGE-L, chrF, and SentenceBERT results, these findings show that task-specific instruction tuning substantially narrows the performance gap and enables smaller base LLMs to surpass large-scale general-purpose LLMs on selected evaluation dimensions.

\begin{rqanswer}{Answer to RQ1}
\approach improves all four base LLMs across all five automatic metrics, with average relative gains of 53.84\% on METEOR and 37.22\% on chrF. Despite using only 7B or 8B base LLMs, its variants achieve the highest overall METEOR and BERTScore, while a single variant, \approach{}-\modelname{Llama-3.1-8B-Instruct}, outperforms every large-scale LLM baseline on at least two metrics. These results demonstrate the consistent effectiveness of \approach across different base LLM families and its ability to make smaller, task-specific models competitive with large-scale general-purpose LLMs.
\end{rqanswer}

\begin{table*}[t]
\centering
\begin{threeparttable}

\caption{Impact of synthesized training-data scale on COBOL section summarization performance on Stack-120.}
\label{tab:synthesized_data_scale}

\small
\renewcommand{\arraystretch}{1.15}
\setlength{\tabcolsep}{5pt}
\sisetup{
    table-number-alignment = center,
    reset-text-series = false,
    text-series-to-math = true,
    reset-text-family = false,
    text-family-to-math = true
}

\begin{tabular}{
    @{}
    c
    c
    c
    S[table-format=2.3]
    S[table-format=2.3]
    S[table-format=2.3]
    S[table-format=2.3]
    S[table-format=2.3]
    @{}
}
\toprule
\textbf{Model Name}
& \textbf{Training Setting}
& \textbf{Training Instances}
& {\textbf{ROUGE-L}}
& {\textbf{METEOR}}
& {\textbf{chrF}}
& {\textbf{BERTScore}}
& {\textbf{SentenceBERT}} \\
\midrule

\multirow{5}{*}{DeepSeek-R1-Distill-Llama-8B}
& Base LLM
& /
& 13.040
& 10.411
& 17.704
& 82.700
& 52.322 \\

\cmidrule(lr){2-8}

& \multirow{4}{*}{\approach}
& 500
& 18.995
& 15.272
& 23.141
& 83.918
& 57.394 \\

&
& 1,000
& 19.617
& 16.895
& 24.502
& 84.279
& 57.395 \\

&
& 2,000
& 19.979
& 18.314
& 25.725
& 84.426
& 58.570 \\

&
& 3,764
& 22.048
& 20.102
& 26.710
& 85.029
& 60.464 \\

\bottomrule
\end{tabular}

\vspace{0.6ex}

\begin{tablenotes}[flushleft]
\footnotesize
\item[] \textit{Note.} ``/'' indicates that training instances are not applicable to the original base LLM.
\end{tablenotes}

\end{threeparttable}
\end{table*}

\subsection{RQ2: Synthesized Data Effectiveness}
\label{sec:rq2}

\noindent\textbf{Motivation.}
Stage~1 of \approach is designed to alleviate training-data scarcity by synthesizing execution-validated COBOL code--summary data. 
We therefore investigate the effectiveness of the synthesized training data and how their scale affects COBOL section summarization performance.

\noindent\textbf{Methodology.}
We use \modelname{DeepSeek-R1-Distill-Llama-8B} as the fixed base LLM. It is a representative and widely adopted open-weight reasoning model at the 8B scale and falls within the locally deployable model range targeted by this study. To isolate the effect of synthesized training-data scale, we first randomly shuffle the 3,764 instances produced in Stage~1 using a fixed random seed. We then take the first 500, 1,000, and 2,000 instances, together with the full set of 3,764 instances, forming four nested training sets. For each scale, we apply the same subsequent supervision-construction and \approach fine-tuning pipeline while keeping the model, training hyperparameters, inference settings, and Stack-120 evaluation set unchanged. The original \modelname{DeepSeek-R1-Distill-Llama-8B} without task-specific fine-tuning serves as the control.

\noindent\textbf{Results.}
As shown in Table~\ref{tab:synthesized_data_scale}, even 500 synthesized training instances improve all five automatic metrics. Compared with the base LLM without task-specific fine-tuning, ROUGE-L, METEOR, and chrF improve by 45.67\%, 46.69\%, and 30.71\%, respectively, while BERTScore and SentenceBERT improve by 1.47\% and 9.69\%. These gains show that the data constructed in Stage~1 provide effective supervision even at a relatively small training scale.

More importantly, all five automatic metrics increase consistently as the synthesized training-data scale grows from 500 to 1,000, 2,000, and 3,764 instances, with the full-data setting achieving the best result on every metric. Scaling from 500 to 3,764 instances yields further relative improvements of 16.07\% on ROUGE-L, 31.63\% on METEOR, 15.42\% on chrF, 1.32\% on BERTScore, and 5.35\% on SentenceBERT. The particularly large additional gain on METEOR indicates that increasing synthesized supervision continues to strengthen the learned summarization behavior.

The benefit also persists at larger training scales. Increasing the synthesized set from 2,000 to 3,764 instances still improves all five metrics, including further gains of 10.36\% on ROUGE-L and 9.76\% on METEOR. The improvement therefore does not rapidly saturate after introducing a small amount of synthesized data; within the evaluated range, increasing the amount of execution-validated synthesized training data continues to provide additional benefits.

\begin{rqanswer}{Answer to RQ2}
The data synthesized in Stage~1 provide effective supervision for COBOL section summarization, and their benefits consistently increase with training-data scale. Scaling from 500 to 3,764 instances improves all five automatic metrics, including further gains of 31.63\% on METEOR, 16.07\% on ROUGE-L, and 15.42\% on chrF. All five metrics continue to improve even when scaling from 2,000 to 3,764 instances, showing that, within the evaluated range, increasing synthesized training data consistently improves summarization performance.
\end{rqanswer}

\begin{table*}[t]
\centering
\begin{threeparttable}

\caption{Impact of training and inference context on COBOL section summarization performance on Stack-120.}
\label{tab:context_ablation}

\small
\renewcommand{\arraystretch}{1.15}
\setlength{\tabcolsep}{3pt}

\begin{tabular*}{\textwidth}{
    @{\extracolsep{\fill}}
    c
    c
    c
    c
    c
    c
    c
    @{}
}
\toprule
\textbf{Model Variant}
& \textbf{Training / Inference Context}
& \textbf{ROUGE-L}
& \textbf{METEOR}
& \textbf{chrF}
& \textbf{BERTScore}
& \textbf{SentenceBERT} \\
\midrule

Base-DeepSeek-R1-Distill-Llama-8B
& Section Code + DD + IE
& 13.040
& 10.411
& 17.704
& 82.700
& 52.322 \\

\approach{}-Code
& Section Code
& 17.334
& 14.925
& 20.962
& 78.336
& 52.931 \\

\approach{}-Code+DD
& Section Code + DD
& 20.039
& 17.289
& 24.482
& 84.461
& 58.117 \\

\approach{}-Code+IE
& Section Code + IE
& 20.600
& 19.087
& 26.479
& 84.623
& 58.306 \\

\approach{}-Full
& Section Code + DD + IE
& 22.048
& 20.102
& 26.710
& 85.029
& 60.464 \\

\bottomrule
\end{tabular*}

\vspace{0.5ex}

\begin{tablenotes}[flushleft]
\footnotesize
\item[] \textit{Note.} DD denotes relevant data division declarations, and IE denotes identifier explanation. For the base LLM, the listed context is used only during inference.
\end{tablenotes}

\end{threeparttable}
\end{table*}

\subsection{RQ3: Expanded Context}
\label{sec:rq3}

\noindent\textbf{Motivation.}
\approach constructs expanded context by augmenting the target section with relevant data division declarations and identifier explanations to address the separation of data and logic. However, it remains necessary to determine whether this additional context improves summarization and how the raw declarations and their natural-language explanations contribute. We therefore examine the effects of relevant data division declarations, identifier explanations, and their combination on COBOL section summarization performance.

\noindent\textbf{Methodology.}
Following RQ2, we use \modelname{DeepSeek-R1-Distill-Llama-8B} as the fixed base LLM and compare four \approach variants that differ only in the context provided during training and inference. \approach{}-Code uses only the target section; \approach{}-Code+DD additionally includes the relevant data division declarations; \approach{}-Code+IE additionally includes the identifier explanation; and \approach{}-Full combines the target section, relevant declarations, and identifier explanation, corresponding to the complete expanded context. For settings containing IE, each identifier explanation is generated from the identifier name and its retrieved data declaration following the standard \approach pipeline. All other factors, including structured rationale supervision, training configurations, and inference settings, remain unchanged. For reference, we also report the original base LLM using the full expanded context at inference time.

\begin{figure}[!t]
\centering
\includegraphics[
    width=\columnwidth,
    keepaspectratio
]{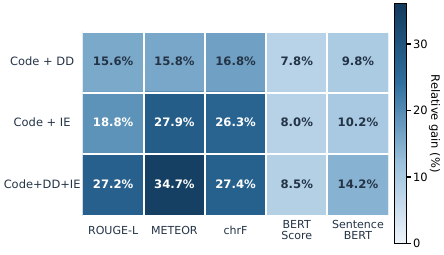}
\caption{Relative gains produced by different expanded-context configurations on Stack-120. All gains are calculated relative to \approach{}-Code, which uses only the target section code during training and inference. DD denotes relevant data division declarations, and IE denotes identifier explanation.}
\label{fig:rq3_context_gain}
\end{figure}

\noindent\textbf{Results.}
Table~\ref{tab:context_ablation} reports the raw scores under different context configurations, while Fig.~\ref{fig:rq3_context_gain} visualizes the relative gains of the expanded-context variants over \approach{}-Code.

Adding relevant data division declarations consistently improves all five metrics. Compared with \approach{}-Code, \approach{}-Code+DD improves ROUGE-L, METEOR, chrF, BERTScore, and SentenceBERT by 15.61\%, 15.84\%, 16.79\%, 7.82\%, and 9.80\%, respectively. This result confirms that local section code alone does not provide sufficient information for constraint-aware summarization. The relevant declarations expose data properties and constraints that are defined outside the procedure division and are therefore unavailable from the target section alone.

The generated identifier explanation provides a more accessible natural-language representation of these declarations. Relative to \approach{}-Code, \approach{}-Code+IE improves ROUGE-L, METEOR, chrF, BERTScore, and SentenceBERT by 18.84\%, 27.89\%, 26.32\%, 8.03\%, and 10.15\%, respectively. It also outperforms \approach{}-Code+DD on every metric, with particularly clear additional gains of 10.40\% on METEOR and 8.16\% on chrF. These results indicate that explicitly explaining identifier roles, data properties, and relevant constraints makes compact and symbolic COBOL declarations easier for the model to use.

The best results are obtained by \approach{}-Full, which combines the target section, relevant data division declarations, and identifier explanation. Compared with \approach{}-Code, it improves ROUGE-L, METEOR, chrF, BERTScore, and SentenceBERT by 27.20\%, 34.69\%, 27.42\%, 8.54\%, and 14.23\%, respectively. It also consistently outperforms \approach{}-Code+IE. This result shows that the original declarations and their natural-language explanations are complementary rather than redundant. The identifier explanation makes the relevant semantics explicit and easier to use, while the original declarations preserve precise source-level constraints that may be simplified or omitted during abstraction.

\begin{rqanswer}{Answer to RQ3}
Expanded context consistently improves COBOL section summarization. Relevant data division declarations provide constraints unavailable in local section code, while the identifier explanation makes these constraints more explicit and easier to use. Combining both representations yields the best performance, improving METEOR by 34.69\% and SentenceBERT by 14.23\% over section code alone. These results confirm that raw declarations and their natural-language explanations provide complementary information.
\end{rqanswer}

\subsection{RQ4: Structured Rationale Supervision}
\label{sec:rq4}

\noindent\textbf{Motivation.}
\approach uses constraint-guided structured rationale supervision to explicitly analyze identifier semantics, data constraints, and section behavior before producing the final summary. However, the performance gains observed in RQ1 may also partly arise from supervised fine-tuning itself rather than from the structured rationale. We therefore investigate whether structured rationale supervision provides additional benefits over direct summary-only supervision.

\noindent\textbf{Methodology.}
Following RQ2 and RQ3, we use \modelname{DeepSeek-R1-Distill-Llama-8B} as the fixed base LLM and compare three settings: the original base LLM without task-specific fine-tuning, Summary-only SFT, and the complete \approach. All settings use the same full expanded context during evaluation. The two fine-tuned settings further use the same training instances, data scale, input context, and hyperparameters, differing only in their supervision targets. Specifically, Summary-only SFT uses only the Phase~4 final summary from each qualified rationale as its supervision target, whereas \approach uses the complete four-phase structured rationale from the same instance, comprising identifier explanation, constraint analysis, logic abstraction, and final summary. Thus, the two fine-tuned settings share exactly the same final-summary supervision, while \approach additionally provides intermediate structured rationale supervision before the final summary. By controlling all other factors, this experiment isolates the effect of structured rationale supervision relative to direct summary supervision.

\begin{table*}[t]
\centering
\begin{threeparttable}

\caption{Effect of structured rationale supervision compared with base and Summary-only SFT settings on Stack-120.}
\label{tab:rationale_ablation}

\small
\renewcommand{\arraystretch}{1.15}
\setlength{\tabcolsep}{3.5pt}
\renewcommand{\tabularxcolumn}[1]{m{#1}}

\begin{tabularx}{\textwidth}{
    >{\hsize=1.90\hsize\linewidth=\hsize
      \centering\arraybackslash}X
    >{\hsize=1.35\hsize\linewidth=\hsize
      \centering\arraybackslash}X
    >{\hsize=0.75\hsize\linewidth=\hsize
      \centering\arraybackslash}X
    >{\hsize=0.70\hsize\linewidth=\hsize
      \centering\arraybackslash}X
    >{\hsize=0.50\hsize\linewidth=\hsize
      \centering\arraybackslash}X
    >{\hsize=0.75\hsize\linewidth=\hsize
      \centering\arraybackslash}X
    >{\hsize=1.05\hsize\linewidth=\hsize
      \centering\arraybackslash}X
}
\toprule
\textbf{Model Name}
& \textbf{Training Setting}
& \mbox{\textbf{ROUGE-L}}
& \textbf{METEOR}
& \textbf{chrF}
& \textbf{BERTScore}
& \mbox{\textbf{SentenceBERT}} \\
\midrule

\multirow{3}{*}{DeepSeek-R1-Distill-Llama-8B}
& Base
& 13.040
& 10.411
& 17.704
& 82.700
& 52.322 \\

& Summary-only SFT
& 15.529
& 13.150
& 20.933
& 82.972
& 53.663 \\

& \approach
& 22.048
& 20.102
& 26.710
& 85.029
& 60.464 \\

\bottomrule
\end{tabularx}

\vspace{0.5ex}

\begin{tablenotes}[flushleft]
\footnotesize
\item[] \textit{Note.} All settings use the same full expanded context during evaluation. The two fine-tuned settings use the same training instances, input context, and hyperparameters.
\end{tablenotes}

\end{threeparttable}
\end{table*}

\begin{figure}[!t]
\centering
\includegraphics[
    width=\columnwidth,
    keepaspectratio
]{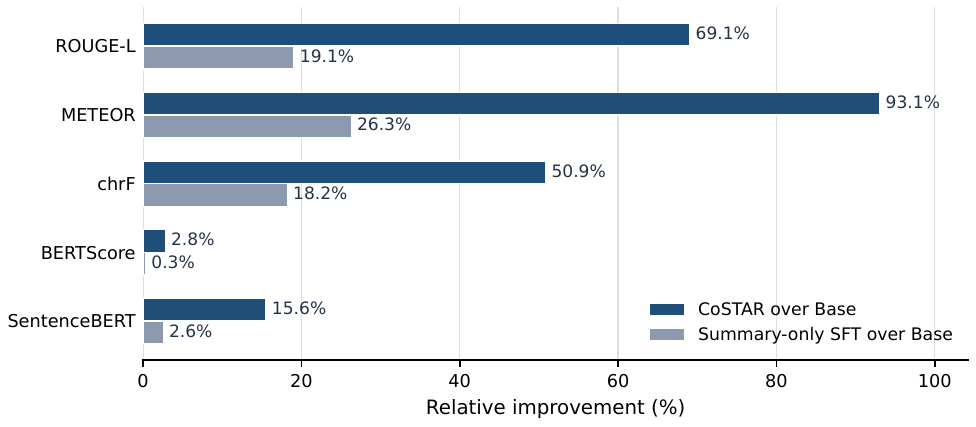}
\caption{Relative improvements of Summary-only SFT and \approach over the same base LLM on Stack-120. All settings use \modelname{DeepSeek-R1-Distill-Llama-8B} and the same full expanded context during evaluation.}
\label{fig:rq4_rationale_gain}
\end{figure}

\noindent\textbf{Results.}
Table~\ref{tab:rationale_ablation} reports the raw scores of the three settings, while Fig.~\ref{fig:rq4_rationale_gain} compares the relative improvements of Summary-only SFT and \approach over the same base LLM.

Summary-only SFT improves the base LLM across all five metrics. Its relative gains reach 19.09\% on ROUGE-L, 26.31\% on METEOR, and 18.24\% on chrF. In contrast, the gains on BERTScore and SentenceBERT are only 0.33\% and 2.56\%, respectively. This pattern suggests that direct summary supervision improves alignment with the reference wording but provides a comparatively limited learning signal for capturing COBOL-specific semantics and data constraints.

\approach produces substantially larger improvements. Relative to the base LLM, it improves ROUGE-L, METEOR, chrF, BERTScore, and SentenceBERT by 69.08\%, 93.08\%, 50.87\%, 2.82\%, and 15.56\%, respectively. Compared directly with Summary-only SFT, \approach further improves the five metrics by 41.98\%, 52.87\%, 27.60\%, 2.48\%, and 12.67\%. Because the two fine-tuned settings use the same base LLM, training samples, data scale, expanded context, and hyperparameters, these additional gains do not arise from a larger model, more training data, or additional input information, but correspond to the difference in supervision.

These results demonstrate that constraint-guided structured rationale supervision provides a more effective task-specific learning signal than direct summary supervision. By explicitly organizing identifier roles, data division constraints, control and data flow, and modernization-relevant behavior, the structured rationale teaches the model how these elements relate before producing the final summary, enabling it to make more effective use of the expanded context.

\begin{rqanswer}{Answer to RQ4}
Summary-only SFT improves the base LLM, but structured rationale supervision provides substantially larger gains. Under the same base LLM, training data, and full expanded context, \approach further improves METEOR by 52.87\% and SentenceBERT by 12.67\% over Summary-only SFT. These results show that the effectiveness of \approach cannot be explained by supervised fine-tuning alone and that constraint-guided structured rationale supervision is a key contributor to its performance.
\end{rqanswer}

\subsection{RQ5: Industrial Applicability}
\label{sec:rq5}

\noindent\textbf{Motivation.}
Legacy-system modernization ultimately involves real-world enterprise COBOL code under confidentiality and local-deployment constraints. Moreover, automatic metrics cannot fully capture migration-relevant information such as data constraints, boundary conditions, and state effects~\cite{hu2022correlating,roy2021reassessing}. We therefore conduct a human evaluation on Industrial-200 to assess the practical applicability of \approach in real-world enterprise modernization scenarios.

\noindent\textbf{Methodology.}
We evaluate \approach in our industrial partner's local environment, where \modelname{Qwen3-235B} is currently deployed to support COBOL code migration. To enable a more controlled same-family comparison, we instantiate \approach with \modelname{Qwen3-8B}, yielding \modelname{CoSTAR-Qwen3-8B}. This setting reduces confounding from cross-family differences and allows us to examine whether task specialization can enable a substantially smaller, locally deployable model to match or surpass the enterprise-deployed LLM. The human evaluation is organized around the participants and evaluation criteria, evaluation procedure, and statistical analysis.

\noindent\textit{Participants and Evaluation Criteria.}
We invite three engineers from our industrial partner to serve as domain experts. Each expert has more than ten years of COBOL development experience and extensive experience with IBM Z mainframes and enterprise legacy systems. Following prior code summarization studies~\cite{su2024distilled,haque2022semantic}, the experts independently assess each generated summary in terms of accuracy, completeness, and conciseness using the four-point rubric shown in Table~\ref{tab:human_evaluation_rubric}. Higher scores indicate better summary quality.

\begin{table*}[t]
\centering
\caption{Human evaluation rubric for COBOL section summaries.}
\label{tab:human_evaluation_rubric}

\small
\renewcommand{\arraystretch}{1.22}
\renewcommand{\tabularxcolumn}[1]{m{#1}}
\setlength{\tabcolsep}{7pt}

\begin{tabularx}{\textwidth}{
    @{}
    >{\centering\arraybackslash}m{0.11\textwidth}
    >{\raggedright\arraybackslash}X
    >{\raggedright\arraybackslash}m{0.47\textwidth}
    @{}
}
\toprule

\multicolumn{1}{c}{\textbf{Metric}} &
\multicolumn{1}{c}{\textbf{Description}} &
\multicolumn{1}{c}{\textbf{Grade Scale}} \\

\midrule

\multicolumn{1}{
    >{\centering\arraybackslash}m{0.11\textwidth}
}{Accuracy} &
Whether the summary is consistent with the target section and relevant data division declarations, without unsupported or hallucinated claims. &
\textbf{4:} Fully accurate, with no unsupported claims.\newline
\textbf{3:} Mostly accurate, with only minor imprecision.\newline
\textbf{2:} Contains notable errors or unsupported claims.\newline
\textbf{1:} Largely incorrect or misleading. \\

\midrule

\multicolumn{1}{
    >{\centering\arraybackslash}m{0.11\textwidth}
}{Completeness} &
Whether the summary covers the principal section behavior and essential migration-relevant information, including data constraints, boundary conditions, and state effects. &
\textbf{4:} Covers all essential information.\newline
\textbf{3:} Covers the main behavior and most essential information.\newline
\textbf{2:} Contains important omissions.\newline
\textbf{1:} Misses the principal behavior or most essential information. \\

\midrule

\multicolumn{1}{
    >{\centering\arraybackslash}m{0.11\textwidth}
}{Conciseness} &
Whether the summary communicates essential information directly, without unnecessary detail, repetition, or poor prioritization. &
\textbf{4:} Focused and succinct.\newline
\textbf{3:} Mostly concise, with minor redundancy.\newline
\textbf{2:} Noticeably verbose or repetitive.\newline
\textbf{1:} Excessively verbose or unfocused. \\

\bottomrule
\end{tabularx}
\end{table*}

\noindent\textit{Evaluation Procedure.}
For each section in Industrial-200, \modelname{CoSTAR-Qwen3-8B} and \modelname{Qwen3-235B} independently generate one summary. Model identities are hidden during evaluation, so the experts do not know which model generated each summary. Each expert independently evaluates each summary with access to the target section and its relevant data division declarations, and rates both model outputs for all 200 sections according to the rubric in Table~\ref{tab:human_evaluation_rubric}. Because Industrial-200 contains confidential business logic, all model inference and human evaluation are conducted within the partner's local environment. No source code or generated summary leaves this environment; only anonymized ratings are returned for statistical analysis.

\noindent\textit{Statistical Analysis.}
For each section, model, and criterion, we first average the ratings assigned by the three experts. The reported means and sample standard deviations are then calculated over the resulting 200 section-level scores. Following prior software engineering studies~\cite{fang2024esale}, we use two-sided Wilcoxon signed-rank tests~\cite{wilcoxon1963critical} to compare the paired section-level scores of the two models. Because accuracy, completeness, and conciseness constitute three related hypotheses, we apply the Holm correction~\cite{holm1979simple} to control the family-wise error rate. Inter-rater agreement is measured using Fleiss' kappa~\cite{fleiss1971measuring} based on the experts' original integer ratings.

\noindent\textbf{Results.}
Table~\ref{tab:industrial_human_evaluation} reports the human evaluation results on Industrial-200.

\begin{table}[t]
\centering
\caption{Human evaluation on Industrial-200. Values are mean (sample standard deviation).}
\label{tab:industrial_human_evaluation}

\footnotesize
\renewcommand{\arraystretch}{1.18}
\renewcommand{\tabularxcolumn}[1]{m{#1}}
\setlength{\tabcolsep}{3.5pt}

\begin{tabularx}{\linewidth}{
    @{}
    >{\centering\arraybackslash}m{0.19\linewidth}
    >{\centering\arraybackslash}X
    >{\centering\arraybackslash}X
    >{\centering\arraybackslash}X
    @{}
}
\toprule
\textbf{Metric} &
\shortstack{\texttt{CoSTAR-}\\\texttt{Qwen3-8B}} &
\texttt{Qwen3-235B} &
\shortstack{\textbf{Holm-adjusted}\\\textbf{$p$-value}} \\
\midrule

Accuracy &
\textbf{3.195} (0.728) &
3.062 (0.730) &
$9.259 \times 10^{-3}$ \\

Completeness &
\textbf{3.083} (0.740) &
2.853 (0.746) &
$3.885 \times 10^{-6}$ \\

Conciseness &
\textbf{3.175} (0.755) &
3.047 (0.752) &
$3.512 \times 10^{-3}$ \\

\bottomrule
\end{tabularx}
\end{table}

\modelname{CoSTAR-Qwen3-8B} achieves higher mean scores than the enterprise-deployed \modelname{Qwen3-235B} on all three criteria. Specifically, it improves accuracy, completeness, and conciseness by 4.35\%, 8.06\%, and 4.21\%, respectively, corresponding to an average relative improvement of 5.54\%. The largest improvement occurs in completeness, indicating that \approach more effectively captures the principal section behavior together with migration-relevant information such as data constraints, boundary conditions, and state effects. The simultaneous improvements in accuracy and conciseness further indicate that this additional information is provided without more unsupported claims or unnecessary verbosity.

All three differences remain statistically significant after Holm correction, with adjusted $p$-values below 0.01. The sample standard deviations range from 0.728 to 0.755 and are similar across the two models, indicating comparable variation in the expert ratings. The overall Fleiss' kappa is 0.763, while the values across the six model--criterion combinations range from 0.732 to 0.794. These results indicate substantial agreement among the three experts and support the reliability of the human evaluation findings.

Overall, despite being built on only an 8B base LLM, \modelname{CoSTAR-Qwen3-8B} significantly outperforms the enterprise-deployed \modelname{Qwen3-235B} across accuracy, completeness, and conciseness. This result shows that task specialization for COBOL section summarization can enable a substantially smaller locally deployable model to achieve better summary quality in real-world enterprise modernization scenarios.

\begin{rqanswer}{Answer to RQ5}
On Industrial-200, \modelname{CoSTAR-Qwen3-8B} outperforms the enterprise-deployed \modelname{Qwen3-235B} in accuracy, completeness, and conciseness by 4.35\%, 8.06\%, and 4.21\%, respectively, with all three differences remaining statistically significant after Holm correction. The overall Fleiss' kappa of 0.763 indicates substantial inter-rater agreement. These findings show that \approach, built on only an 8B base LLM, can outperform a substantially larger enterprise-deployed LLM under real-world confidentiality and local-deployment constraints.
\end{rqanswer}
\section{Threats to Validity}
\label{sec:threats}

\noindent\textbf{Internal Threats.}
Internal threats mainly concern the reliability of training supervision construction and potential data leakage. Although compilation and test execution filter incorrect generated programs, finite test suites cannot guarantee complete semantic correctness. We therefore filter low-quality tasks and employ a judge LLM~\cite{wu2025can} to control the quality of teacher-generated structured rationales and reference summaries. Nevertheless, automated verification and LLM-based evaluation cannot completely eliminate noise or hallucinations. The synthesized training set, Stack-120, and Industrial-200 are constructed independently to avoid direct overlap; however, the non-transparent pretraining corpora of base LLMs prevent us from fully excluding prior exposure to public COBOL code or task descriptions.

\noindent\textbf{External Threats.}
External threats concern the generalizability of our findings. Although our data synthesis methodology is language- and granularity-agnostic, it is instantiated through CodeFlowBench and evaluated only on COBOL section-level summarization. Applying it to other languages, granularities, or tasks may require adjusted prompts and validation criteria. Moreover, variations in compiler dialects, coding conventions, and application domains may limit generalization to broader COBOL legacy systems. Although \approach does not require a shared backbone, we use the same base LLM for both modules to control backbone variation. Future work will explore heterogeneous backbone combinations, broader COBOL environments, and other low-resource legacy languages.

\noindent\textbf{Construct Threats.}
Construct threats concern whether our evaluation methodology accurately reflects COBOL section summarization quality. Since conventional metrics may not fully capture COBOL-specific semantics such as field constraints and state updates, we combine complementary automatic metrics with expert evaluation on Industrial-200. Human evaluation may introduce subjectivity; therefore, three enterprise engineers with more than ten years of COBOL experience independently rated the summaries using a shared four-point rubric, achieving substantial agreement (Fleiss' kappa = 0.763). Reference quality is another potential threat. Although judge-based quality control is applied, teacher-generated structured rationales and their Phase~4 reference summaries may still contain inaccuracies. For Stack-120, each reference summary is independently reviewed by three enterprise COBOL experts, with disagreements resolved through discussion.
\section{Conclusion}
\label{sec:conclusion}

In this paper, we present \approach, an integrated framework for COBOL section summarization in legacy system modernization that unifies execution-validated data synthesis with constraint-aware model training and inference. The synthesis stage transforms general-purpose programming tasks into execution-validated COBOL code-summary data, and experiments further confirm that these synthesized data provide effective training supervision for section summarization. Across four 7B or 8B base LLMs, \approach improves all five automatic metrics, with average relative gains of 53.84\% on METEOR and 37.22\% on chrF. Ablation studies further confirm the effectiveness of expanded context and structured rationale supervision. On real-world enterprise COBOL systems, \modelname{CoSTAR-Qwen3-8B} significantly outperforms the enterprise-deployed \modelname{Qwen3-235B} in accuracy, completeness, and conciseness. Overall, these results show that execution-validated data synthesis and constraint-aware reasoning can enable small, locally deployable LLMs to effectively support COBOL program understanding and modernization. Future work will extend our study to more industrial systems, COBOL dialects and compiler environments, program granularities, and other low-resource programming languages.

\bibliographystyle{IEEEtran}
\bibliography{IEEEabrv,GOUDA}

\end{document}